\documentclass[12pt]{article}
\usepackage[margin=1.25in]{geometry}

\usepackage{graphicx}

\usepackage{scrtime} 
\usepackage{wrapfig}
\usepackage{rotating}
\usepackage{multirow}
\usepackage{enumitem}
\usepackage[normalem]{ulem}
\usepackage{appendix}
\usepackage{bm} 

\usepackage{xcolor}
\definecolor{grape}{RGB}{128, 0, 225}
\definecolor{aqua}{RGB}{0, 128, 225}
\definecolor{strawberry}{RGB}{225, 0, 128}
\definecolor{links}{RGB}{54, 115, 41}
\definecolor{extlinks}{RGB}{6, 71, 137}
\definecolor{bggrey}{RGB}{247, 247, 247}
\definecolor{knitcmt}{RGB}{172,150,174}
\definecolor{knitkwd}{RGB}{186,91,102}
\definecolor{knitstr}{RGB}{54,128,201}
\definecolor{knitid}{RGB}{88,88,88}
\definecolor{jrsiblue}{RGB}{62, 86, 159}

\usepackage{xspace}

\usepackage{authblk}

\usepackage{amsmath, amssymb}

\usepackage{tikz}
\usepackage{animate}

\usepackage{listings}
\lstdefinestyle{customc}{
  backgroundcolor=\color{bggrey},
  belowcaptionskip=1\baselineskip,
  breaklines=true,
  xleftmargin=\parindent,
  language=C,
  showstringspaces=false,
  basicstyle=\ttfamily,
  keywordstyle=\bfseries\color{green!50!black},
  commentstyle=\itshape\color{grape!},
  identifierstyle=\color{blue},
  stringstyle=\color{red},
}
\lstdefinestyle{customR}{
  backgroundcolor=\color{bggrey},
  belowcaptionskip=1\baselineskip,
  breaklines=true,
  xleftmargin=\parindent,
  language=R,
  showstringspaces=false,
  basicstyle=\ttfamily,
  keywordstyle=\bfseries\color{knitkwd},
  commentstyle=\itshape\color{knitcmt},
  identifierstyle=\color{knitid},
  stringstyle=\color{knitstr},
}

\usepackage{cite}
\usepackage{cleveref}

\newcommand{\betacos}{\beta_{\cos}}
\newcommand{\betattf}{\beta_{\rm tt}}

\DeclareMathOperator{\osc}{osc}
\newcommand{\betahigh}{\beta_{\rm high}}
\newcommand{\betalow}{\beta_{\rm low}}

\newcommand{\ps}{p_{\rm{s}}}

\newcommand{\pow}{\mathcal{P}}
\newcommand{\R}{\mathcal{R}}
\newcommand{\meanbeta}{\langle \beta \rangle}
\newcommand{\eps}{\varepsilon}
\newcommand{\avgrho}{\rho_0}

\newcommand{\ie}{{\sl i.e., }}
\newcommand{\eg}{\emph{e.g.,} }

\newcommand{\xpp}{\texttt{xppaut}\xspace}
\newcommand{\auto}{\texttt{AUTO}\xspace}

\newcommand{\supp}{\href{https://royalsocietypublishing.org/action/downloadSupplement?doi=10.1098\%2Frsif.2019.0202\&file=rsif20190202supp1.pdf}{\color{extlinks}electronic supplementary material}\xspace}

\definecolor{Rlimegreen}{RGB}{50, 205, 50}
\definecolor{Rdarkgreen}{RGB}{0, 100, 0}
\definecolor{Rdarkorchid1}{RGB}{191, 62, 255}
\definecolor{Rdarkorchid4}{RGB}{104, 34, 139}

\newcommand{\eref}[1]{equation~\ref{eqn:#1}}

\newcommand{\fref}[1]{figure~\ref{fig:#1}}
\newcommand{\Fref}[1]{Figure~\ref{fig:#1}}

\newcommand{\tref}[1]{table~\ref{tab:#1}}

\crefname{app}{appendix}{appendices}

\newcommand{\slab}[1]{\label{sec:#1}}

\newcommand{\sref}[1]{§\ref{sec:#1}}

\newcommand{\suppfref}[1]{\href{run:rsif20190202supp1.pdf}{{\color{extlinks}\supp, figure~\ref*{fig:#1}}}}
\newcommand{\suppsref}[1]{\href{run:rsif20190202supp1.pdf}{{\color{extlinks}\supp, §\ref*{sec:#1}}}}

\newcommand{\msfref}[1]{\href{run:forceshape.pdf}{{\color{extlinks}figure~\ref*{fig:#1}}}}
\newcommand{\msmethods}{\href{run:forceshape.pdf}{{\color{extlinks}Methods \xspace}}}

\newcommand \dbyd[2]  { \frac{\mathrm d{#1}}{\mathrm d{#2}}   }   

\global\let\oldnewlabel\newlabel
\gdef\newlabel#1#2{\newlabelxx{#1}#2}
\gdef\newlabelxx#1#2#3#4#5#6{\oldnewlabel{#1}{{#2}{#3}}}
\let\newlabel\oldnewlabel
\numberwithin{equation}{section}

\title{Invariant predictions of epidemic patterns from radically different forms of seasonal forcing
}

\author[a]{Irena Papst*}
\author[b,c]{David J.D. Earn}
\affil[a]{Center for Applied Mathematics, 657 Frank H.T. Rhodes Hall, Cornell University, Ithaca, NY, 14853, United States}
\affil[b]{Department of Mathematics and Statistics, McMaster University, 1280 Main Street West, Hamilton, ON L8S 4K1, Canada}
\affil[c]{M. G. DeGroote Institute for Infectious
Disease Research, 1280 Main Street West, Hamilton, ON L8S 4K1, Canada}
\date{\today, \thistime}

\begin{document}

\maketitle

\begin{center}
*Corresponding author (email: ip98@cornell.edu)
\end{center}

\vspace{2em}


\vspace{2em}

\noindent \textbf{Cite this article:} Papst I, Earn DJD. 2019 Invariant predictions of epidemic patterns from radically different forms of seasonal forcing. \emph{J. R. Soc. Interface} \textbf{16}: 20190202. \url{http://dx.doi.org/10.1098/rsif.2019.0202}

\newpage


\begin{abstract}

Seasonal variation in environmental variables, and in rates of contact among individuals, are fundamental drivers of infectious disease dynamics. Unlike most periodically-forced physical systems, for which the precise pattern of forcing is typically known, underlying patterns of seasonal variation in transmission rates can be estimated approximately at best, and only the period of forcing is accurately known. Yet solutions of epidemic models depend strongly on the forcing function, so dynamical predictions---such as changes in epidemic patterns that can be induced by demographic transitions or mass vaccination---are always subject to the objection that the underlying patterns of seasonality are poorly specified.  Here, we demonstrate that the key bifurcations of the standard epidemic model are invariant to the shape of seasonal forcing if the amplitude of forcing is appropriately adjusted.  Consequently, analyses applicable to real disease dynamics can be conducted with a smooth, idealized sinusoidal forcing function, and qualitative changes in epidemic patterns can be predicted without precise knowledge of the underlying forcing pattern.  We find similar invariance in a seasonally forced predator-prey model, and conjecture that this phenomenon---and the associated robustness of predictions---might be a feature of many other periodically forced dynamical systems.
\end{abstract}

\section{Introduction}

Periodic forcing of infectious disease transmission arises from a number of sources, including seasonal weather changes\cite{ShamKohn09,He+13b}, annual cycles in birth rates\cite{HeEarn07,Mart+14}, and school terms\cite{LondYork73,Earn+00,Ston+07,Earn+12}.  While the period of forcing is always one year, the pattern of forcing depends strongly on the source (\eg nearly sinusoidal for weather or births, but sharp and asymmetric for school terms\cite{Sche84,Earn+00}) and the forcing function is never known exactly.

The detailed structure of the seasonal forcing function strongly affects the resulting pattern of disease incidence.  For example, \fref{forcedts} compares solutions of the standard seasonally forced \emph{susceptible-infectious-recovered} (SIR) epidemic model \cite{AndeMay91,BrauCast01} (see \sref{methods}) with different forcing patterns (school terms or a sinusoid).  In each case, the basic reproduction number ($\R_0$) and mean infectious period ($1/\gamma$) are the same, and the amplitude of seasonality ($\alpha$) has been chosen to ensure that the system displays a strictly biennial cycle.  The differences between the time series in \fref{forcedts} emphasize that the solution of a model can match the details of an observed temporal pattern of prevalence only if the pattern of forcing is known (or estimated accurately).

However, in situations in which we are primarily interested in the length of recurrent cycles of epidemics---or whether or not the recurrent epidemic pattern exhibits dynamical chaos---it is not clear that knowing the detailed forcing pattern is critical.  Indeed, it was noted in ref.~\citen{Earn+00} (endnote 13) that in a parameter regime relevant to measles dynamics, qualitatively similar bifurcation structure \emph{can} be obtained with different seasonal forcing functions.  In particular, the structure of the bifurcation diagram that facilitates predictions of observed dynamical transitions (bifurcation parameter $\R_0$ with fixed seasonal forcing amplitude $\alpha$) is qualitatively similar for school term forcing with amplitude estimated from data ($\alpha=0.25$) or sinusoidal forcing with much lower amplitude ($\alpha=0.08$).

Here, for a wide range of forcing functions, we examine the quantitative and qualitative bifurcation structure of the SIR model (below) and a seasonally forced predator-prey model (in \suppsref{supppredprey}).  Revealing invariance of dynamical transitions in these systems deepens our understanding of epidemic and predator-prey dynamics, and suggests a more general type of invariance in the dynamical structure of forced nonlinear oscillators.

\section{Methods}
\slab{methods}
\subsection{Seasonally forced SIR model}

The standard seasonally forced SIR model for a closed population can be expressed as a system of three ordinary differential equations,
\begin{subequations}
\label{eqn:SIR}
  \begin{align}
    \dbyd{S}{t} &= \mu - \beta(t) SI -\mu S\,,
    \label{subeqn:SIRs} \\
    \noalign{\vspace{5pt}}
    \dbyd{I}{t} &= \beta(t) SI - (\gamma + \mu) I\,, \label{subeqn:SIRi} \\
    \noalign{\vspace{5pt}}
    \dbyd{R}{t} &= \gamma I - \mu R \,, \label{subeqn:SIRr}
  \end{align}
\end{subequations}
where $S$, $I$ and $R$ denote the proportions of the population that are susceptible, infectious and recovered from the infection, $\mu$ is the \emph{per capita} birth and death rate, $\gamma$ is the recovery rate, and $\beta(t)$ is the time-dependent transmission rate.  The first two equations above do not depend on the variable $R$, and $R=1-S-I$, so the third equation is not required to specify the dynamics.

In most previous work, the transmission rate $\beta(t)$ has been taken to vary sinusoidally for simplicity\cite{HethLevi89,OlseScha90,Heth00,Dush+04},
\begin{equation}
  \label{eqn:betacos}
  \betacos(t) = \meanbeta [1 + \alpha \cos (2 \pi t)],
\end{equation}
where $\meanbeta$ is the mean transmission rate, $\alpha$ is the amplitude of seasonality, and time is measured in units of the forcing period (one year).  For childhood diseases such as measles, mumps, rubella or whooping cough, the primary source of seasonal forcing is typically school terms\cite{LondYork73}.  Consequently, a substantial body of work has employed a forcing function that is high in ``term-time'' and low otherwise\cite{Sche84,BolkGren93,Earn+00,BaucEarn03,BaucEarn03b,HempEarn15},
\begin{equation}
  \label{eqn:betattf-1}
  \betattf(t) = \begin{cases} \betahigh &\mbox{school days}, \\
    \betalow & \mbox{other days},
  \end{cases}
\end{equation}
where $\betahigh$ and $\betalow$ are constant transmission rates, with
$\betahigh > \betalow$.  If $\ps$ is the proportion of the year that children spend in school then
\begin{equation}
\meanbeta = \ps \betahigh + (1 - \ps) \betalow,
\end{equation}
so if we define the amplitude of seasonality to be
\begin{equation}
\alpha = \frac{1}{2} \left ( \frac{\betahigh - \betalow}{\meanbeta} \right ),
\end{equation}
and denote the term-time ``oscillation'' function to be
\begin{equation}
  \label{eqn:betattf-2prep}
 \osc_{\rm tt}(t) = \begin{cases} 2(1-\ps) &\mbox{school days}, \\
    -2 \ps & \mbox{other days},
  \end{cases}
\end{equation}
we can rewrite \eref{betattf-1} in terms of the two parameters $\meanbeta$ and $\alpha$,
\begin{equation}
  \label{eqn:betattf-2}
 \betattf(t) = \meanbeta [1 + \alpha \osc_{\rm tt}(t)]\,.
\end{equation}
The amplitude of forcing, $\alpha$, is constrained to lie in the range\cite{BaucEarn03b}
\begin{equation}\label{eqn:alphaconstraint}
0 \le \alpha \le \frac{1}{2\ps} \, ,
\end{equation}
to ensure that the forcing function is always non-negative. Regardless of the form of periodic forcing, the basic reproduction number (the average number of secondary cases per primary case in a wholly susceptible population) is\cite{MaMa06}
 \begin{equation}
   \label{eqn:R0seas}
   \R_0 = \frac{\meanbeta}{\gamma+\mu} \,.
 \end{equation}

\subsection{Family of forcing functions}
\label{sec:famff}

In order to better understand how the dynamical structure of the SIR model depends on the pattern of seasonality in transmission, we constructed a continuous family of forcing functions that connects term-time and sinusoidal forcing.  The family is parameterized with a \emph{shape parameter} $p$, such that the forcing follows the school term schedule for $p=-1$, is square and symmetric for $p=0$, and is sinusoidal for $p=1$.  The forcing function, which we denote $\osc_p(t)$, is symmetric for $p\ge0$ and asymmetric for $p<0$ (more time above than below the mean).  For $p>1$, the sine wave is squashed and in the limit $p\to\infty$ approaches two delta function impulses (up at integral times of $t=0, 1, 2, ...$ years, and down at half-integral times of  $t=0.5, 1.5, 2.5, ...$ years).  Several members of this family of forcing functions are plotted in \fref{ff-fam}; details of the construction can be found in \suppsref{defin-family-forc}.

\subsection{Stroboscopic map}

Perturbative analysis of periodic orbits is usually simplified by considering a Poincar\'{e} map, which discretely samples continuous trajectories each time they pass through a particular hypersurface in the state space \cite[\S8.7]{Stro18}.  For periodically forced systems, the natural Poincar\'{e} map is the \textit{stroboscopic map}, which samples trajectories once per forcing period (\textit{cf.} Example 8.7.2 in \cite{Stro18}).  Our bifurcation analysis is based on the one-year stroboscopic map for the SIR system \eqref{eqn:SIR}, so $n$-year periodic attractors are represented by $n$ discrete points.

\subsection{Continuation of bifurcations}
\label{sec:contbif}

It is rarely possible to find explicit analytical expressions for bifurcation points as functions of model parameters.  Instead, the conditions that define a given bifurcation are typically solved numerically and then ``continued'' in parameter space (\ie starting from one bifurcation point, a curve in parameter space is computed on which the bifurcation conditions are satisfied to some level of precision\cite{Kuzn04}).  We used standard open-source software (XPPAUT\cite{Erme02}) to continue bifurcation points in one and two model parameters.

The one-parameter bifurcation diagrams shown in \fref{R0bd-alpha-adj} were constructed by finding equilibria and cycles of the annual stroboscopic map of the seasonally forced SIR model, and then continuing those points as functions of $\R_0$ (details in \suppsref{comp-bif-diags}).

To find the sinusoidal forcing amplitude that yields a bifurcation diagram (\fref{R0bd-alpha-adj}, bottom panel) that is similar to that of the term-time forced model (\fref{R0bd-alpha-adj}, top panel), we proceeded as follows.  Starting from the stable period doubling (PD) bifurcation point highlighted with a square in the term-time bifurcation diagram (\fref{R0bd-alpha-adj}, top panel, $\R_0=15.12$), we continued the bifurcation in the two-dimensional $(p,\alpha)$ parameter space.  At each point on the resulting curve, shown in \fref{compalphap}, the value of $\R_0$ for the PD remains the same.  Consequently, this curve can be thought of as a function, $\alpha(p)$, which specifies the forcing amplitude ($\alpha$) that yields a PD at the same $\R_0$ value for any given shape of forcing pattern ($p$).  In particular, the second point marked with a square in \fref{compalphap} shows that the forcing amplitude that yields a PD at $\R_0=15.12$ with sinusoidal forcing ($p=1$) is $\alpha=0.1$, the amplitude used to make the bifurcation diagram shown in the bottom panel of \fref{R0bd-alpha-adj}.

\section{Results and Discussion}

\subsection{Quantitative invariance}

\Fref{R0bd-alpha-adj} displays unexpected similarity of bifurcation diagrams for SIR models with radically different forcing functions.
The close correspondence between the positions of bifurcations in the top and bottom panels of \fref{R0bd-alpha-adj} is surprising given how the diagrams were constructed.  We first created the top panel (term-time forced SIR model bifurcation diagram as a function of $\R_0$).  We then made the middle panel, using all the same parameter values but with sinusoidal rather than term-time forcing.  To make the bottom panel, we used sinusoidal forcing again but adjusted the forcing amplitude so that the position ($\R_0$ value) of the principal period doubling bifurcation matched the position of the same bifurcation in the top panel ($\R_0=15.12$; see \sref{methods}).  Remarkably, with no further adjustments, the five fold bifurcations\cite{Kuzn04} listed in \tref{bif-lineup} match to similar precision.  These fold bifurcations occur in different parts of the parameter space on disconnected branches; yet their positions are conserved for a continuous family of forcing functions between term-time and sinusoidal (see \fref{bif-lineup}, \tref{bif-lineup}, and \supp), suggesting that the global dynamical structure of the model is determined by a property of the forcing function that is independent of its shape.

\subsection{Qualitative invariance}

Some aspects of the invariance are qualitative rather than quantitative.
While the fold bifurcations at the left of the five ``fold branches'' in the $\R_0$ bifurcation diagrams (\fref{R0bd-alpha-adj}) appear to be invariant provided the position of the principal period doubling is conserved, the folds at the right edges of these branches do not line up precisely.  Thus, the observed quantitative invariance is restricted to the ``births'' of these branches and not their ``deaths''.  Nevertheless, the main qualitative structure is preserved.

\subsection{Unknown source of invariance}

What the bifurcation-conserving property might be is not clear.  Preliminary work indicated that fixing the average spectral power in the forcing function might be sufficient to conserve the bifurcation structure\cite{BaucEarn03}, but detailed analysis shows this to be false (see \suppfref{compalphap-full}).  With the hope of identifying another spectral feature that might be associated with invariance of bifurcations, we also considered the shape of the full power spectrum of the forcing function.  Unfortunately, as the pattern of forcing is smoothly changed from term-time to sinusoidal while keeping the bifurcation structure constant (see \suppfref{powspecs}), the full spectrum varies dramatically and no specific spectral feature underlying the observed bifurcation invariance was apparent to us.

\subsection{Summary of results}
\label{sec:results}

\Fref{bif-lineup} and \tref{bif-lineup} summarize our main results.  Our continuation of the stable period doubling (PD) bifurcation (\fref{compalphap}) determines the forcing amplitude that yields the same $\R_0$ value for the PD for each member of the family of forcing functions, including term-time ($p=-1$, $\alpha=0.25$) and sinusoidal ($p=1$, $\alpha=0.1$).  This fact is indicated in \fref{bif-lineup} by the vertical alignment of the black squares (each square occurs at $\R_0=15.12$).  The filled circles of each colour are also aligned vertically in \fref{bif-lineup}, showing that as $\R_0$ increases the positions of all the folds that create attractors with periods longer than two years are invariant with respect to the forcing pattern.  \tref{bif-lineup} shows that these ``birth folds'' are invariant to similar precision as the PD as it is continued.  The open circles and squares in \fref{bif-lineup} show that the positions of other bifurcations---the corresponding ``death folds'' and some intermediate PDs that appear in the longest period fold branches as $p$ is increased---are not invariant to the shape of the forcing function.

\section{Conclusion}

The bifurcation invariance that we have observed in the seasonally forced SIR model has significant practical importance, since the form of forcing in real infectious disease systems is never known precisely.  

From a theoretical perspective, valuable rigorous analyses (\eg proving the co-existence of multiple attractors\cite{SchwSmit83} or the existence of chaos\cite{GlenPerr97}) can be performed more easily if sinusoidal forcing is assumed; given the bifurcation invariance, such analyses are likely less dependent on the pattern of forcing than might previously have been thought.

From an applied perspective, predicting transitions in epidemic patterns \cite{Earn+00,BaucEarn03,KrylEarn13,HempEarn15} depends on estimates of the pattern of forcing \cite{FineClar82,Hook+11,Kryl11,deJo14}, which can change in both amplitude and shape over decades or centuries \cite{Kryl11,HempEarn15}; the bifurcation invariance helps explain why transition analyses have worked so well \cite{Earn+00,BaucEarn03,Kryl11,HempEarn15}, in spite of crude estimates of transmission forcing.

Why such bifurcation invariance occurs, and the extent to which it can be generalized to other periodically forced dynamical systems\cite{Stro18,KaplGlas95,Winf01}, remains to be seen.  The SIR model is one of the very simplest nonlinear systems, having only a single quadratic nonlinearity.  As a first step towards generalization, we have found that the type of invariance exhibited by the SIR model is also displayed by a periodically forced predator-prey model (\suppsref{supppredprey}).  It seems likely that similar invariances occur in many other forced dynamical systems.  If a property can be identified that produces the observed invariance, this may lead to important general insights about periodically forced dynamical systems.

In the meantime, dynamical phenomena observed in idealized population models employing sinuoidal forcing\cite{SchwSmit83, GlenPerr97, Dush+04, Rina+93, GragRina95, Sche+97, Kuzn+92} can reasonably be assumed to be relevant to epidemics and ecosystems in the real world, and epidemiological transition analyses \cite{Earn+00,BaucEarn03,Kryl11,KrylEarn13,HempEarn15} can be conducted with confidence using imperfectly estimated transmission forcing.

\subsection*{Data accessibility}

Code required to replicate the results presented in this paper is available in \suppsref{supp-code}.

\subsection*{Authors' contributions}
IP and DJDE designed and performed research, analysed data, and wrote the paper; IP wrote computer programs and performed simulations. Both authors gave final approval for publication and agree to be held accountable for the work performed therein.

\subsection*{Competing interests}
We declare we have no competing interests.

\subsection*{Funding}
IP was supported by NSERC postgraduate scholarships (CGS-M, PGS-D) and an Ontario Graduate Scholarship.  DJDE was supported by NSERC.

\subsection*{Acknowledgements}
We are grateful to the Earn and Ellner research groups, Sigal Balshine, Ben Bolker, Stephen Ellner and Steven Strogatz, for discussions and comments.

\bibliography{forceshape}

\newpage
\begin{table*}[h!]
\centering
\caption{\textbf{Invariance of fold bifurcations{\rm\cite{Kuzn04}} at different $\bm{\R_0}$ values when the principal period doubling (PD) bifurcation at $\bm{\R_0=15.12}$ is matched.}
\textmd{Fold ($n$) refers to a fold bifurcation that gives rise to a period $n$ attractor (an $n$-year epidemic cycle) as $\R_0$ is increased.  The PD does not occur at precisely the same $\R_0$ for each $(p,\alpha)$ pair due to slight inaccuracies of the numerical continuation software\cite{Erme02,DoedOlde11}.
The relative difference refers to $\max[(x-x_{\rm tt})/x_{\rm tt}]$, where $x$ is the value of $\R_0$ at the bifurcation of interest and $x_{\rm tt}$ is its value for term-time forcing ($p=-1$).
All the data in this table are plotted in \fref{bif-lineup}.}}
\label{tab:bif-lineup}
\begin{tabular}{|r|c|r|r|r|r|r|r|}
\hline
\multicolumn{2}{|c|}{\bfseries Forcing Function}&\multicolumn{6}{c|}{\bfseries $\bm{\mathcal{R}_0}$ Bifurcation Point}\tabularnewline
\cline{1-8}
\multicolumn{1}{|c|}{$p$}&\multicolumn{1}{c|}{$\alpha$}&\multicolumn{1}{c|}{PD (2)}&\multicolumn{1}{c|}{Fold (3)}&\multicolumn{1}{c|}{Fold (4)}&\multicolumn{1}{c|}{Fold (5)}&\multicolumn{1}{c|}{Fold (6)}&\multicolumn{1}{c|}{Fold (7)}\tabularnewline
\hline
$-1.00$&$0.2500$&$15.1199$&$8.8991$&$6.2374$&$4.9755$&$4.2623$&$3.8094$\tabularnewline
$-0.50$&$0.1012$&$15.1254$&$8.9014$&$6.2363$&$4.9713$&$4.2554$&$3.7995$\tabularnewline
$ 0.00$&$0.0782$&$15.1146$&$8.8946$&$6.2247$&$4.9554$&$4.2357$&$3.7758$\tabularnewline
$ 0.25$&$0.0839$&$15.1248$&$8.8992$&$6.2292$&$4.9599$&$4.2401$&$3.7800$\tabularnewline
$ 1.00$&$0.1000$&$15.1234$&$8.8976$&$6.2269$&$4.9571$&$4.2370$&$3.7767$\tabularnewline
$ 2.00$&$0.1182$&$15.1240$&$8.8972$&$6.2261$&$4.9558$&$4.2354$&$3.7749$\tabularnewline \hline \hline
\multicolumn{2}{|c|}{\bfseries Relative Difference}&$ 0.0004$&$0.0005$&$0.0020$&$0.0040$&$0.0063$&$0.0090$\tabularnewline
\hline
\end{tabular}
\end{table*}

\newpage

\begin{figure*}
\begin{center}
\scalebox{0.9}{
\includegraphics{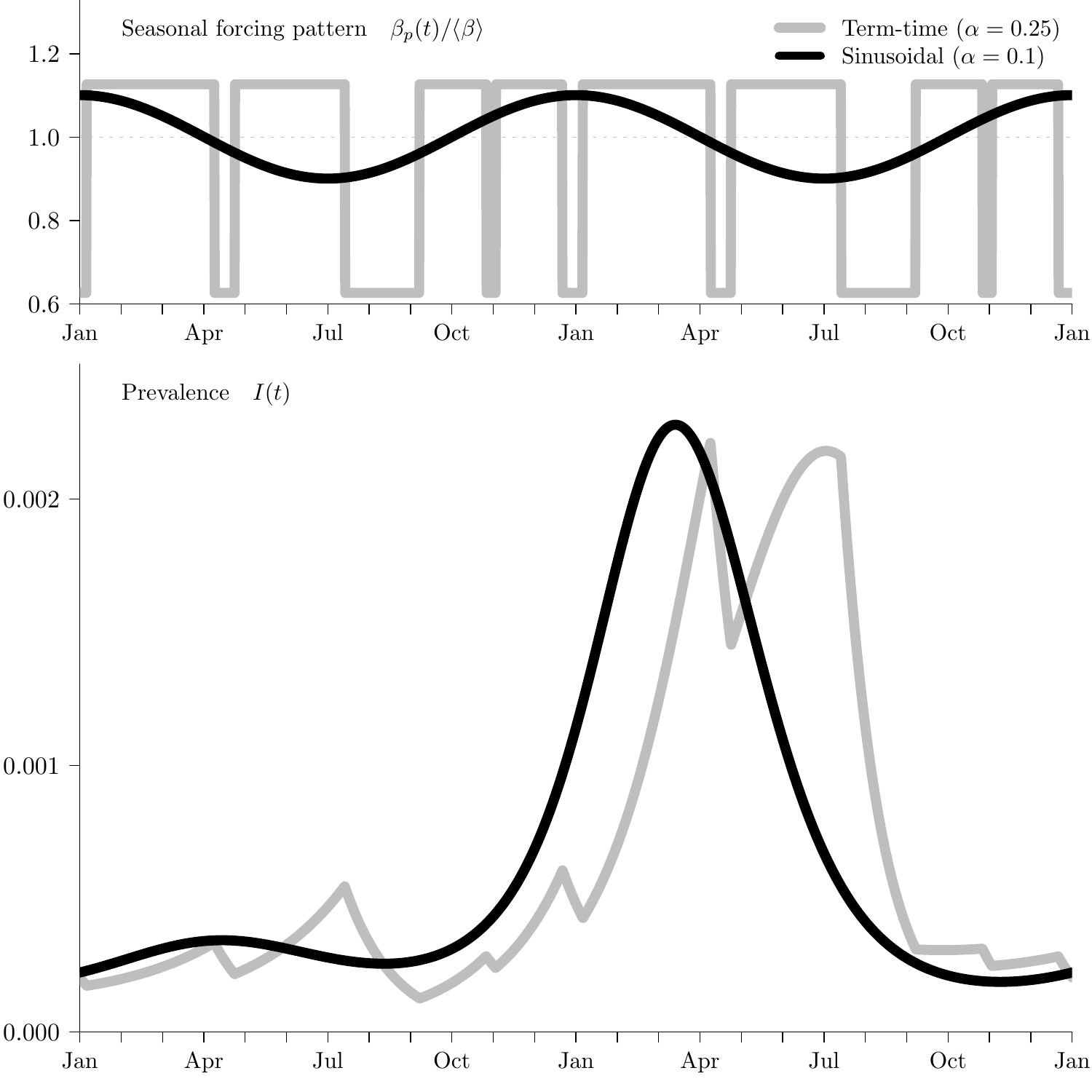}
}
\end{center}
\caption{\textbf{The effects of different seasonal forcing patterns on predicted prevalence time series.} \emph{Top panel:} Patterns of seasonality, $\beta_p(t)/\meanbeta=1+\alpha\osc_p(t)$, for term-time ($p=-1$) and sinusoidal ($p=1$) forcing; see \sref{methods}~and \suppsref{defin-family-forc} for the definition of $\osc_p(t)$.  \emph{Bottom panel:} Prevalence time series, $I(t)$, for solutions of the SIR model (\eref{SIR}) with term-time and sinusoidal forcing.
Parameter values\cite{KrylEarn13}: $\R_0=17$, $1/\gamma=13$ days, $\mu=0.02/\text{year}$, $\alpha=0.25$ (term-time), $\alpha=0.1$ (sinusoidal).
Initial conditions: $(S_0,I_0)=(0.97, 0.03)$.  The prevalence time series are plotted after a 101 year transient.}
\label{fig:forcedts}
\end{figure*}

\begin{figure*}
\begin{center}
\scalebox{0.7}{
\includegraphics{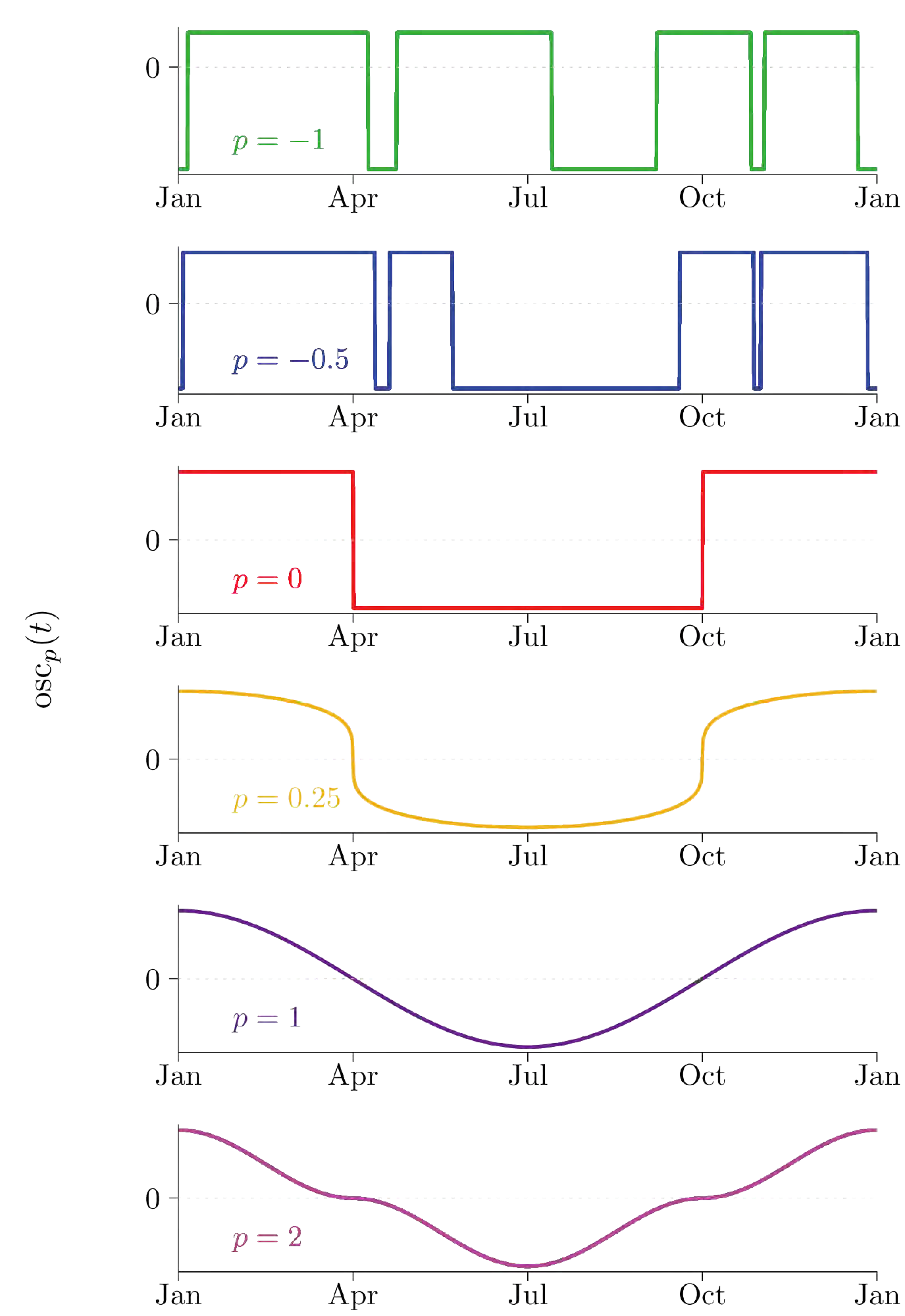}
}
\end{center}
\caption{\textbf{Several members of the family of forcing functions, $\osc_p(t)$, described in \sref{famff}.} \emph{Panels, from top to bottom:} $p=-1$ (term-time forcing), $p=-0.5$, $p=0$ (square wave forcing), $p=0.25$, $p=1$ (sinusoidal forcing), $p=2$. These shape parameter values correspond to those used in \fref{bif-lineup}. Details of the construction are given in \suppsref{defin-family-forc}.}
\label{fig:ff-fam}
\end{figure*}

\begin{figure*}[h!]
\scalebox{0.8}{
\includegraphics{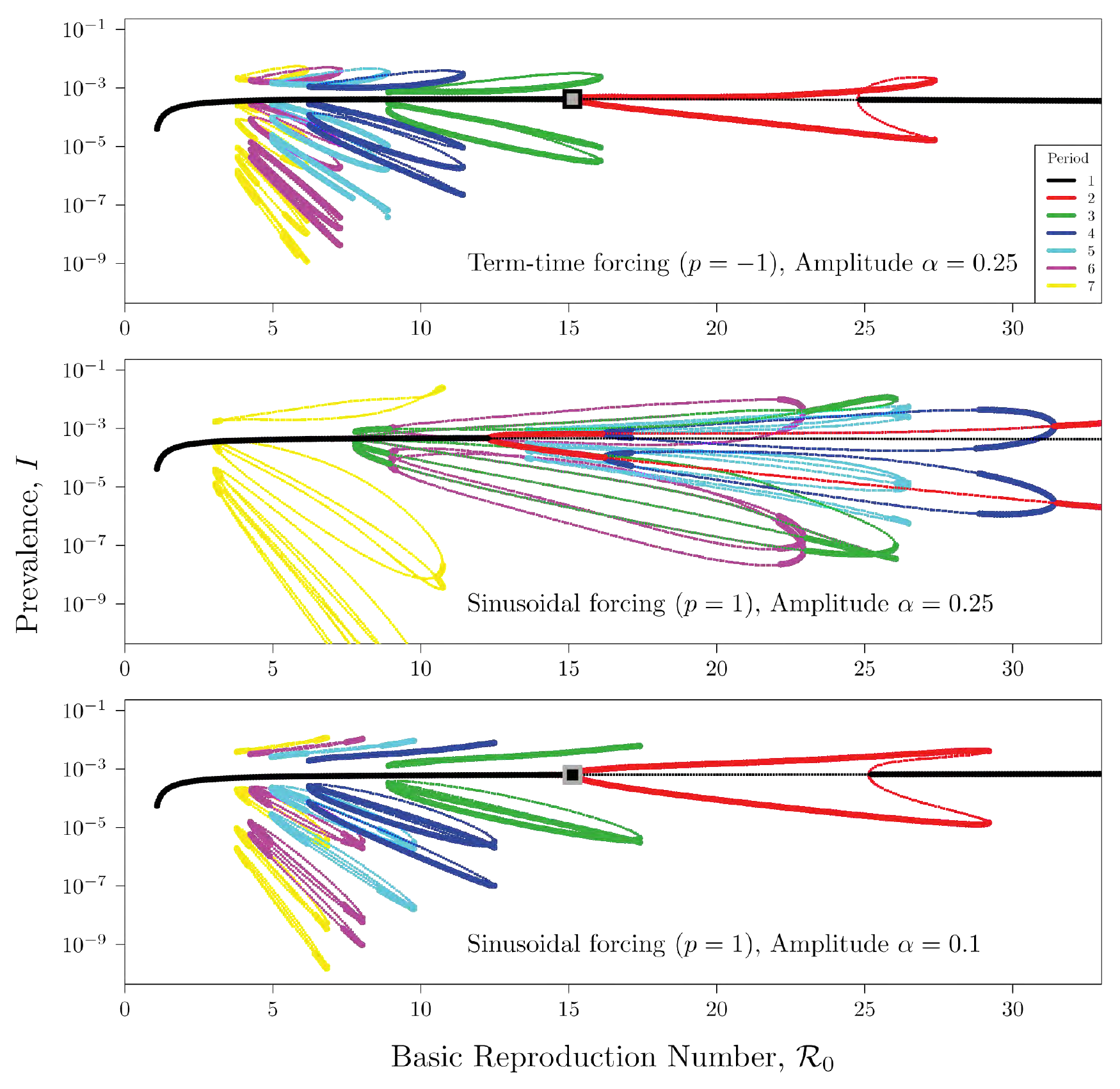}
}

\caption{\textbf{$\bm{\R_0}$ bifurcation diagrams for the annual stroboscopic map of the seasonally forced SIR model (\eref{SIR}) with different patterns and amplitudes of forcing.}  In the top two panels, the forcing pattern is different but the associated amplitudes are the same.  In the bottom two panels, the forcing pattern is the same but the amplitudes are different.  The fixed parameter values are $\mu=0.02/\text{year}$ and $1/\gamma=13$ days (corresponding to measles\cite{KrylEarn13}).  The values of parameters that vary are indicated in each panel.  Thick lines show stable periodic solutions (attractors) and thin lines show unstable periodic solutions (repellors).  At each $\R_0$, the number of points of a given colour indicates the period of the associated attractor or repellor.  The qualitative similarity of the top and bottom panels shows that different forcing patterns can yield the same bifurcation structure (for different forcing amplitudes).  A precise quantitative correspondence of bifurcation points is demonstrated in \fref{bif-lineup} and \tref{bif-lineup}.  The points highlighted with squares in the top and bottom panels (at $\R_0=15.12$) correspond to the similarly highlighted points in the two-dimensional bifurcation diagram in \fref{compalphap}.}
\label{fig:R0bd-alpha-adj} \end{figure*}

\begin{figure*}[h!]
  \centering
  \scalebox{0.85}{
  \includegraphics{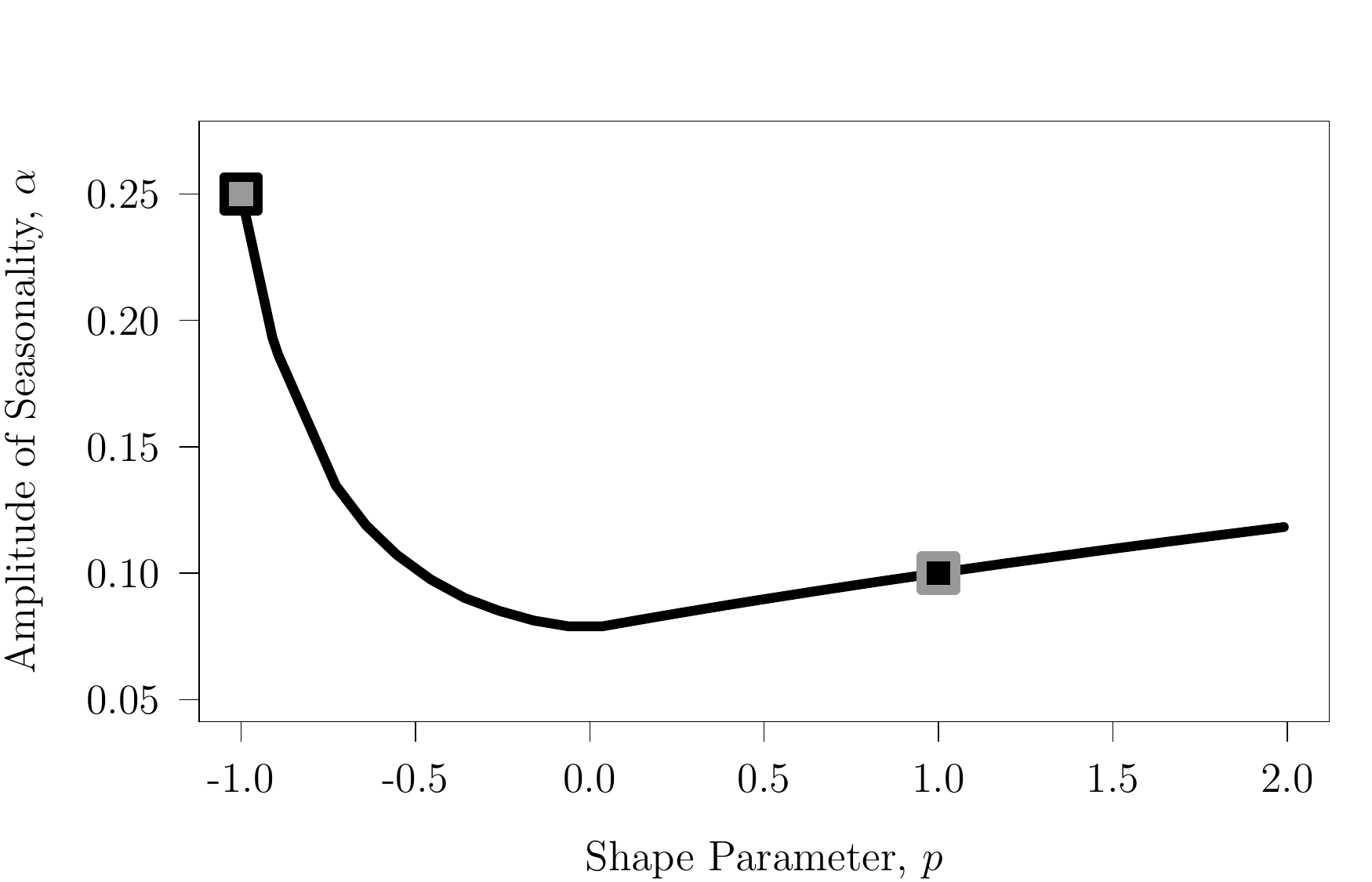}
  }
  \caption{\textbf{Continuation of the stable period doubling (PD) bifurcation in the two-dimensional $\bm{(p,\alpha)}$ parameter plane} (see \sref{methods}~and \suppsref{twoparbifdiags}).  The continuation was initiated at term-time forcing ($p=-1$) with the amplitude estimated from data\cite{Earn+00,BaucEarn03,KrylEarn13} ($\alpha=0.25$) and extends beyond sinusoidal forcing ($p=1$).  The resulting function, $\alpha(p)$, shows how the amplitude of seasonality ($\alpha$) must change as the forcing pattern ($p$) is changed, if we wish to fix the values of all the other model parameters (in particular, $\R_0=15.12$).  The points highlighted with squares correspond to the identically highlighted points in the top and bottom panels of \fref{R0bd-alpha-adj}.}
\label{fig:compalphap}
\end{figure*}

\begin{figure*}[h!]
  \centering
  \scalebox{0.85}{
  \includegraphics{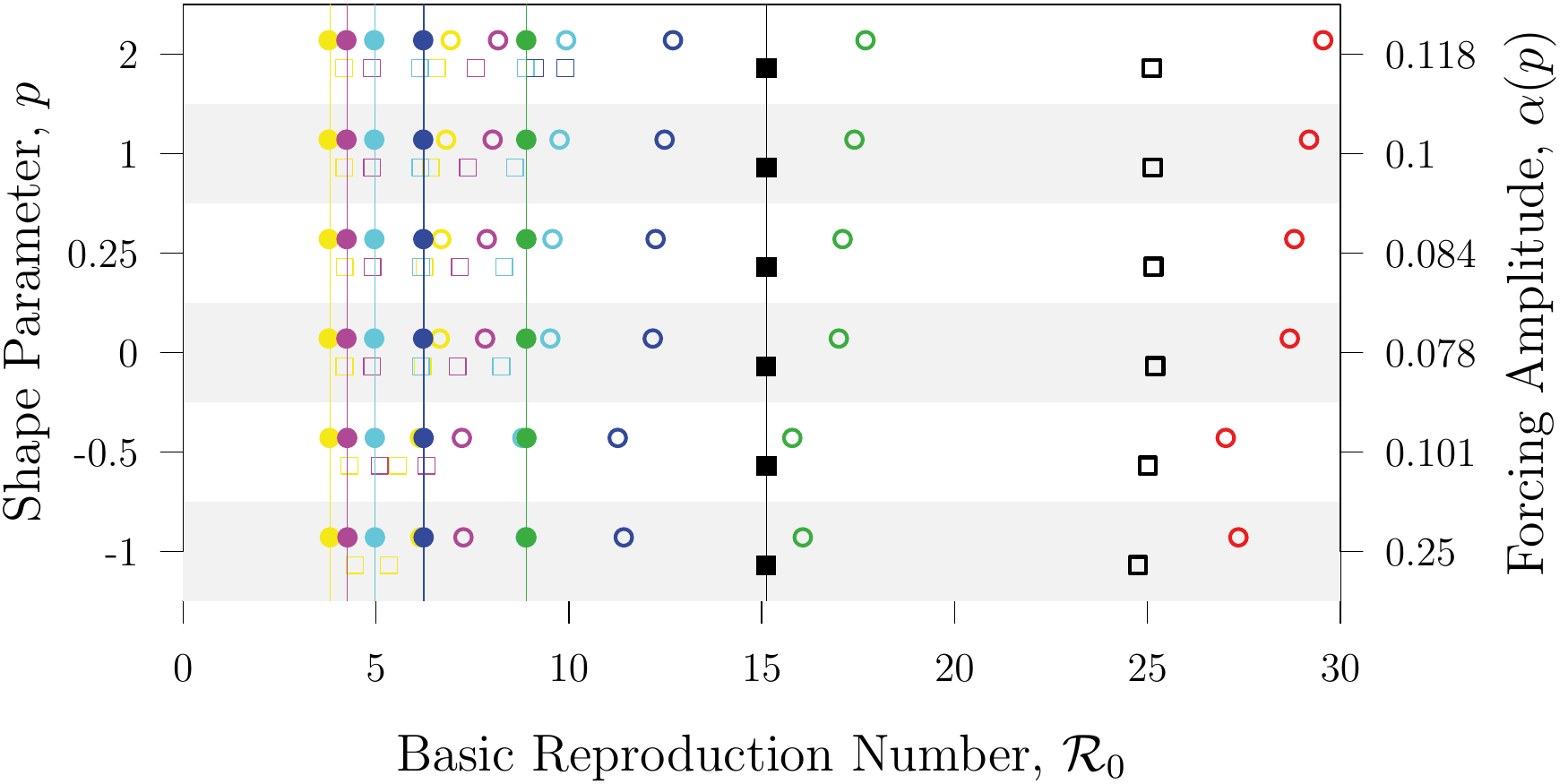}
  }
  \caption{\textbf{Graphical representation of bifurcation invariance in the seasonally forced SIR model (\eref{SIR})}.  For six forcing patterns ($p$, left vertical axis) and amplitudes determined by the function shown in \fref{compalphap} ($\alpha(p)$, right vertical axis), the values of $\R_0$ (horizontal axis) at which bifurcations occur are indicated (with colours that correspond to those used in \fref{R0bd-alpha-adj}).  Period doubling (PD) bifurcations are shown with squares and fold bifurcations are shown with circles.  The PD that is fixed by construction is shown with solid black squares.  Folds that turn out to be invariant are marked with solid circles (these are all of the ``birth folds'').  Other non-invariant bifurcations (``death folds'' and intermediate PDs) are marked with open symbols.  The full $\R_0$ bifurcation diagrams for $p=-1$ (term-time forcing) and $p=1$ (sinusoidal forcing) are shown in the top and bottom panels of \fref{R0bd-alpha-adj}; the full bifurcation diagrams associated with each of the six forcing patterns are shown in \suppsref{suppR0bifdiags}.}
  \label{fig:bif-lineup}
\end{figure*}

\end{document}